\documentclass[aps,prl,twocolumn,showpacs,superscriptaddress]{revtex4}

\begin{document}


\title{On an Application of Relative Entropy}


\author{Dmitry V. Khmelev}
\email{D.Khmelev@newton.cam.ac.uk}
\affiliation{Isaac Newton Institute for Mathematical Sciences, 20
  Clarkson Rd, Cambridge, CB3 0EH, U.K.}  
\affiliation{Heriot-Watt
  University, Edinburgh, U.K. and Moscow State University, Russia}

\author{William J. Teahan}
\email{wjt@informatics.bangor.ac.uk}
\affiliation{University of Wales, Bangor, Dean Street, Bangor, LL57 1UT, U.K.}

\date{\today}

\begin{abstract}
  We show that in problems of authorship attribution and other
  linguistic applications, a Markov Chains approach is a more attractive
  technique than Lempel-Ziv based compression.
\end{abstract}


\maketitle

We wish to point out a number of inaccurate and misleading statements
that Benedetto {\em et al.} make in their paper titled ``Language
Trees and Zipping''\cite{Bene:2002}. First, they claim the technique
they used for construction of a language tree does not make use of any
a-priori information about the alphabet, but it does, both in the
alphabet chosen (Unicode) and in the set of languages they chose to
experiment with; second, they propound Lempel-Ziv (LZ, {\em gzip})
compression as being applicable to DNA analysis, where the usefulness
of LZ is quite doubtful; third, in practice their definition of
relative entropy and distance can yield negative values; fourth, the
classification performance of the method they use is significantly
worse than other entropy-based methods as has been noted in prior
work; and fifth, the classification speed is significantly worse as
well, which shows that its ``potentiality'' is questionable. We
elaborate on each of these points in more detail in the subsequent
paragraphs.

Notice that the ``Language Tree''(LT) diagram~\cite{Bene:2002} does not
include the Russian language (Slavic family of Indo-European family of
languages; 288 million speakers). Our computations show that once
Russian is included, it does not cluster with the other members of
the Slavic group. Obviously, certain Cyrillic alphabet based languages
were left out of the study~\cite{Bene:2002}, which ``improves''
results significantly and shows that a-priori information about
the alphabet is being taken advantage of to achieve the results outlined in
paper~\cite{Bene:2002}.

The LZ compressor makes few assumptions about the input string, but in
practice, we do have a-priori information that we can take advantage
of. Biologists widely use an amino acid {\em substitution matrix} (PAM250
or BLOSUM62) in search for {\em similar} biological
sequences~\cite{gusfield}. It is not at all clear how a substitution
matrix could be implemented with the LZ algorithm.  That is why
compression is not widely used for DNA analysis, although first trials
for its application go back to 1990~\cite{gusfield}.

The quantity $S_{\mathcal A\mathcal B}$~\cite{Bene:2002} defined as
``relative entropy'' in (1) and redefined as ``distance'' in (2) can
take negative values. Negative values indeed appeared in our study
which showed that the ``LT''\cite{Bene:2002} reflects
significantly the structure of Unicode or vice versa, and its
relevance to language classification should be supported additionally.

A traditional definition and estimates for (relative) entropy via $n$th
order Markov Chain {\em on letters}~\cite{Shannon,Yaglom,Kukush:2001}
always lead to a proper positive number. Markov Chains are also
traditional in text entropy analysis~\cite{Shannon,Yaglom},
compression~\cite{PPM}, authorship and subject
attribution~\cite{Teahan:2000,Khmelev:2000}.  In~\cite{Kukush:2001},
the classification performance of compression programs was compared
with the Markov Chain approach~\cite{Khmelev:2000}. 82 authors of
large enough texts ($\ge 10^5$ characters) were chosen. Afterwards 82
one-per-author texts were held out and used for control purposes.  The
classification algorithm~\cite{Kukush:2001} had to
determine the author of each control text among 82 alternatives. The
corresponding numbers of exact guesses for 15 compression programs and
Markov Chains are presented in the following list~\cite{Kukush:2001}:

Program(number of guesses): 7zip(39), arj(46), bsa(44), compress(12), dmc(36),
gzip(50), ha(47), huff(10), lzari(17), ppmd5(46), rar(58), rarw({\bf
  71}), rk(52); Markov Chain approach (see~\cite{Khmelev:2000}) {\bf
  69} guesses.



Clearly, {\em gzip} is significantly outperformed by other compression
algorithms and the first order Markov chain model~\cite{Khmelev:2000}.
Notice also that in practical implementations, the {\em gzip}-based
approach~\cite{Bene:2002} is significantly slower than the first order Markov
chains method~\cite{Khmelev:2000}.

To sum up, in natural language processing (and, perhaps, in other
fields) the $n$th order Markov chain
models~\cite{Khmelev:2000,Teahan:2000} are more appropriate than
an LZ-approach~\cite{Bene:2002}.

\bibliography{cmc}

\end{document}